\documentclass[aps,prl,floats,twocolumn]{revtex4}

\usepackage{amsmath}
\usepackage{amssymb}
\usepackage{graphicx}
\usepackage{textcomp}
\usepackage{dcolumn}

\newcommand{\be}{\begin{equation}}
\newcommand{\ee}{\end{equation}}
\newcommand{\cms}{Co$_2$MnSi}
\newcommand{\cmsl}{Co$_2$MnSi }

\begin{document}

\title{Structural stability, magnetic and electronic properties of \cms(001)/MgO heterostructures: A density functional theory study}

\author{Bj{\"o}rn H{\"u}lsen}
\affiliation{Fritz-Haber-Institut der Max-Planck-Gesellschaft, Faradayweg 4-6, D-14195 Berlin, Germany}

\author{Peter Kratzer}
\email[]{peter.kratzer@uni-duisburg-essen.de}
\affiliation{Fachbereich Physik, Universit{\"a}t Duisburg-Essen, Lotharstr. 1, D-47048 Duisburg, Germany}

\author{Matthias Scheffler}
\affiliation{Fritz-Haber-Institut der Max-Planck-Gesellschaft, Faradayweg 4-6, D-14195 Berlin, Germany}

\date{\today}

\begin{abstract}
A computational study of the epitaxial \cms(001)/MgO(001) interface relevant to tunneling magnetoresistive (TMR) devices is presented.
Employing {\em ab initio} atomistic  thermodynamics, we show that the Co- or MnSi-planes of bulk-terminated \cmsl form stable interfaces, while pure Si or pure Mn termination requires non-equilibrium conditions. Except for the pure Mn interface, the half-metallic property of bulk \cmsl is disrupted by interface bands. Even so, at  homogeneous Mn or Co interfaces these bands contribute little to the minority-spin conductance through an MgO barrier, and hence such terminations could perform strongly in TMR devices.
\end{abstract}

\maketitle

In recent years, tunneling magneto-resistance (TMR) devices, consisting of two ferromagnetic electrodes separated by a tunneling barrier in the form of a thin oxide layer, have been subject to intense research.
This is due to the use of TMRs in non-volatile magnetic random access memories (MRAMs), which are expected to play an increasing role in future electronics.
The TMR device allows for electric read-out of magnetically stored information, as the tunneling resistance is strongly dependent on   the relative orientation of the magnetization in both electrodes (parallel or anti-parallel). 
The difference between the electrical conductances for these two orientations, divided by the smaller of the two, called the TMR ratio, could possibly be huge if the material used for one or both electrodes is a ferromagnetic half-metal. 
In such a material, the Fermi energy $E_F$ lies in a gap for the electrons of one spin, while the density of states (DOS) is metallic for the electrons of the other spin. 
Examples of present interest are the Heusler alloys Co$_2$MnSi and Co$_2$FeSi (and variants thereof), materials that are predicted by 
calculations\cite{Ishida1995b, Galanakis2002b, Picozzi2002} 
to be bulk half-metals.
In a real TMR device, spin-flip processes related to electronic interface states could invalidate the advantages of half-metallic electrodes.\cite{Mavropoulos2005} Even so, spin-selective tunneling may help to operate these devices with high TMR ratio.

Experimentally, progress towards devices with large TMR ratio has been achieved by using MgO tunneling barriers. The success of this barrier material was traced back to selection rules based on band symmetry that may greatly enhance the ratio of the spin-polarized currents.\cite{Mathon2001,Butler2001}
By combining an MgO barrier with \cmsl electrodes,
TMR ratios as high as 192\%~\cite{Ishikawa2006} 
and even 753\%~\cite{Tsunegi2008} 
have been achieved.
However, the underlying materials issues limiting the achievable TMR ratio have largely remained elusive, mostly due to the unknown structure of the ferromagnet-oxide interface.

Here, we employ density functional theory (DFT) calculations to provide comprehensive information about the structural stability, the magnetic and the electronic properties of the Co$_2$MnSi(001)/MgO(001) interface. 
The highly accurate all-electron full-potential linearized augmented plane wave (FP-LAPW) method implemented in the Wien2k code \cite{Schwarz2003} is used to calculate total energies and electronic structure of the interfaces. The generalized gradient approximation (PBE96 \cite{Perdew1996}) to the exchange-correlation functional has been adopted because the GGA gives better results than the LDA (regarding lattice constants and bulk moduli) for Co$_2$MnSi
~\cite{Picozzi2002, Hashemifar2005}. 
For bulk \cmsl, the GGA-PBE calculations give us a gap in the minority-spin DOS of 0.84eV, and a spin gap (between the top of the minority valence band and $E_F$) of 0.33eV~\cite{Huelsen2009}, in excellent agreement with previous calculations\cite{Picozzi2002}.
The muffin-tin radii were set to $R_{MT}\text{(Mg)} = R_{MT}\text{(O)} = 1.8$~bohr, $R_{MT}\text{(Co)} = R_{MT}\text{(Si)} = 2.0$~bohr and $R_{MT}\text{(Mn)} = 2.1$~bohr. A {\bf k}-point mesh of $16 \times 16 \times 2$ and in the interstitial region a plane wave expansion cutoff of 21~Ry was used. All internal atomic coordinates were relaxed until the forces on the nuclei were smaller than 2~mRy/bohr.

The TMR elements are modeled in supercell geometry, using 3 layers of MgO and 11 or 13 layers of the Heusler alloy, depending on the termination at the interface. 
The supercell is chosen symmetric with respect to a mirror plane at the central layer of the MgO barrier, with two identical interfaces. 
Epitaxial growth of the heterostructure is assumed. 
We use a $(1 \times 1)$ unit cell with the calculated lattice constant of \cms, 5.629{\AA} parallel to the interface, and allow for a tetragonal distortion of the thin MgO film. 
(The bulk lattice constants of  \cmsl and MgO differ by $~7$\% in PBE-GGA). 
The length of the supercell in (001) direction is optimized for each interface in a series of calculations, thus allowing for the bond distance at the interface to adjust.

First, we study the thermodynamic stability of the \cms/MgO interface.
Four different terminations of \cms(001) (by CoCo, MnSi, MnMn or SiSi planes), each in combination with four different registries relative to  MgO(001) (O top, Mg top, bridge, or hollow site) have been taken into consideration. By comparing the DFT total energies, we find that for the CoCo, MnSi and MnMn terminated \cmsl the O top site is lower in energy than any other registry, and thus stable. For these three cases, the Mg top site is metastable, while the bridge site is unstable, and the atoms relax to the O top site instead. For the SiSi termination, the Mg top site is found to be the most stable position.

\begin{figure}
\includegraphics[width=7.1cm]{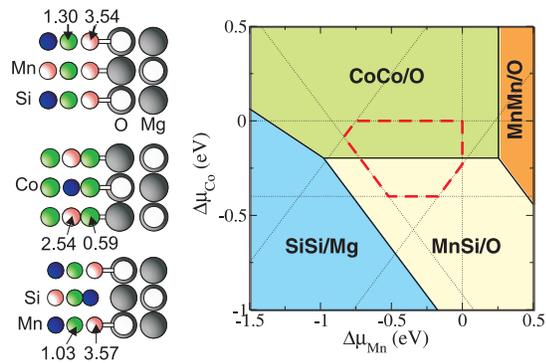}
\caption{\label{fig_phasediagram} (Color online) Left: Ball-and-stick models of the relaxed atomic structures of the MnMn/O interface (top), 
the CoCo/O interface (middle), and the MnSi/O interface (bottom). 
The numbers give the spin magnetic moment of the atoms in units of the Bohr magneton. Right: Phase diagram for the epitaxial \cms(001)/MgO(001) interface. The colored regions correspond to different interface terminations being stable under the conditions described by the chemical potentials $\Delta\mu_i = \mu_i - g_i$, where $g_i$ ($i=$ Co, Mn) is the total energy of one unit cell of the stable bulk phase of the elements (hcp Co and $\alpha$-Mn). The polygon bounded by the red dashed  lines indicates the region accessible in thermodynamic equilibrium with the bulk phases of Mn, Co, Mn$_3$Si or Co$_3$Si. }
\end{figure}

In order to compare the stability of {\it different} terminations, we use the method of {\it ab initio} atomistic thermodynamics \cite{Weinert1987,Scheffler1988}.
The interface energy, normalized to the area $A$, is
\be
\gamma(T,p) = \frac{1}{2A} \left[ \Delta G(N_i,T,p) - \sum_i \Delta N_i \mu_i(T,p) \right].
\label{eq_gamma}
\ee
Here, $\Delta G$ is the difference in Gibbs free energy between the heterostructure and appropriate amounts of the two bulk materials forming the interface. $\Delta N_i$ ($i$ = Co, Mn, Si) counts the number of atoms in which the supercell deviates from bulk composition, and $\mu_i$ is the chemical potential of element $i$.
Since the interface energy is a free-energy {\em difference}, it is often a good approximation to calculate $\gamma$ from the differences of total energies obtained from DFT calculations. Differences in the other contributions to $G$ (due to, e.g., the free energy of lattice vibrations) largely cancel, and are therefore neglected\cite{footnote}.
For the thermodynamic analysis, it is assumed that the interfaces are in thermodynamic equilibrium with the bulk Heusler alloy. This allows to eliminate one chemical potential via the relation 
$ 2 \mu_{\text{Co}} + \mu_{\text{Mn}} + \mu_{\text{Si}} = g_{\text{\cms}}$.
Here, $g_{\text{\cms}}$ is the Gibbs free energy of \cmsl per formula unit. 
The interface energy thus depends on two independent chemical potentials, which we chose to be $\mu_{\text{Co}}$ and  $\mu_{\text{Mn}}$, as in previous studies. 
The interface phase diagram (Fig.~\ref{fig_phasediagram}) displays the termination with the minimal interface energy for every pair $(\mu_{\text{Mn}}, \mu_{\text{Co}})$. 

We conclude from the phase diagram shown in Fig.~\ref{fig_phasediagram} that both the termination of the Heusler electrode by a Co plane, sitting on top of oxygen, as well as termination by a MnSi plane, with both species sitting on top of oxygen, are thermodynamically stable, depending on the preparation conditions of the \cms(001)/MgO(001) interface. The other two terminations considered are not accessible within the limits set by thermodynamic equilibrium with other phases. 
However, they could possibly be fabricated using out-of-equilibrium conditions, and therefore will be considered further.
Our DFT calculations show sizeable structural relaxations at the MnSi-interface (Fig.~\ref{fig_phasediagram}, lower left). 
An Mn-O bond with a length of 2.27{\AA} is formed, while the Si atoms relax towards the Heusler alloy and shorten their bonds to the Co atoms in the next-nearest crystal plane (sublayer). 
In contrast, the Co- and Mn-terminated interfaces remain planar, with  Co-O and Mn-O bond length of 2.09{\AA} and 2.40{\AA}, respectively (Fig.~~\ref{fig_phasediagram}, middle and upper left). 
These results agree with previous DFT studies using a pseudopotential approach.\cite{Miura2007} 

\begin{table}
\begin{ruledtabular}
\begin{tabular}{lD{.}{.}{2.2}D{.}{.}{2.2}D{.}{.}{2.2}D{.}{.}{2.2}}
 & \multicolumn{1}{c}{CoCo/O} & \multicolumn{1}{c}{MnSi/O} & \multicolumn{1}{c}{MnMn/O} & \multicolumn{1}{c}{SiSi/Mg} \\
\hline
$P_{int}$              & 0.67    &  -0.01   & 1.00   & -0.25   \\
$E_{\text{VBO}}$~(V)       & 2.24  & 2.45  & 2.69  & 0.03  \\
\end{tabular}
\caption{\label{tab_mtj_properties} 
Spin polarization $P_{\text{int}}$  of the density of states at the Fermi level, calculated by projecting onto the orbitals of the interface and sublayer atoms, for four terminations of the \cms/MgO(100) interface. $E_{\text{VBO}}$ is the band offset between the valence band of MgO and the minority-spin valence band of \cms.
}
\end{ruledtabular}
\end{table}

Magnetic moments of interface atoms are generally very sensitive to atomic structure. 
Our calculations show that only the  spin magnetic moments at the interface and in the first sublayer of the Heusler alloy (cf. Fig. \ref{fig_phasediagram}) deviate significantly from the bulk moments, 
$m_{\text{Co}}=1.07\mu_B$, $m_{\text{Mn}}=2.91\mu_B$, and $m_{\text{Si}}=-0.04\mu_B$. 
Comparison of the calculated moments  in Fig. \ref{fig_phasediagram} to experimental values can be helpful in drawing conclusions about the interface chemistry.  
Measurements using X-ray magnetic circular dichroism on two \cmsl films of different thickness pointed to an increase of the magnetic moments at the interface to MgO.\cite{Saito2008}
Agreement with this experiment is closest if  we assume a MnMn/O interface (despite the Co-rich film composition reported in Ref.~\onlinecite{Saito2008}), as only this structure shows enhanced Co moments.

\begin{figure}
\centering
\includegraphics[width=7.1cm]{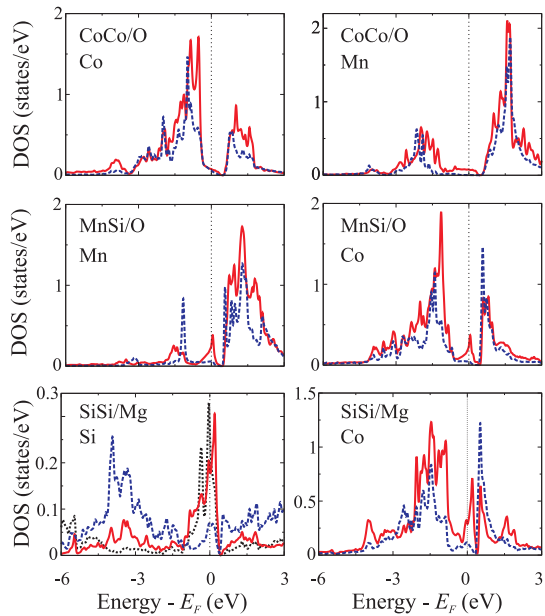}
\caption{(color online) The orbital-resolved DOS of the atoms at the interface (left column) and in the first sublayer (right column) of the Heusler alloy for the three non half-metallic heterostructures CoCo/O, MnSi/O and SiSi/Mg (rows). The second label in each sub-plot denotes the species that is projected out. Red (full) lines show the projected DOS for the 'out-of-plane' orbitals ($d_{3z^2 - r^2} + d_{xz} + d_{yz}$ for Co and Mn, $p_z$ for Si). Blue (dashed) lines show the projected DOS for the 'in-plane' orbitals ($d_{x^2 - y^2} + d_{xy}$ and $p_x + p_y$). The black (dotted) line is a projection onto the Si-$s$ orbital.}
\label{fig_mtj_dos_orbitale}
\end{figure}

The interplay between the atomic and the electronic structure of the interfaces is best analyzed by inspecting the 
orbital-resolved densities of states (DOS).
In Fig. \ref{fig_mtj_dos_orbitale}, the orbital-resolved DOS in the minority spin channel is shown for  the three non-half-metallic interfaces. It is observed that the CoCo/O interface has only a small DOS at $E_F$ which involves both in-plane and out-of-plane Co orbitals, as well as out-of-plane orbitals of the sublayer Mn atoms. The MnSi/O interface shows a sharp peak of the DOS at $E_F$ due to an interface state. The orbital analysis reveals that it arises from the out-of-plane orbitals of the sublayer Co atoms that hybridize with out-of-plane Mn orbitals in the interface layer, similar to the surface states studied previously\cite{Hashemifar2005}.
In case of the SiSi/Mg interface, the half-metallic gap is completely filled with interface states originating from Si orbitals. 
These interface states may drastically reduce the spin polarization of the  DOS at the Fermi level, $P_{\text{int}}$ in Table~\ref{tab_mtj_properties}.
We find that the spin polarization breaks down (and even changes sign) for the MnSi/O and SiSi/Mg interfaces. Only the MnMn/O interface is half-metallic, while for the CoCo/O interface, a rather high degree of spin polarization, $P_{\text{int}}=67$\%, is retained despite the interface state.
Similar to the \cms(001) {\em surface}\cite{Hashemifar2005}, the strong hybridization of $d_{xz}$ and $d_{yz}$ orbitals of both first-layer Mn and second-layer Co atoms prevents the splitting-off of an interface state in this case.

The results for the valence band offsets $E_{\text{VBO}}$ reported in Table~\ref{tab_mtj_properties} are calculated from the self-consistent Kohn-Sham potential (see Ref.~\onlinecite{Godby1994} for the approximations involved). 
The electronic core levels are used to define a common reference energy\cite{Massidda1987}. All interfaces show a sizable valence band offset, with the exception of the SiSi/Mg interface, where the two valence bands are almost perfectly aligned. Hence, the SiSi/Mg interface is clearly unsuitable for TMR devices. Using as input $E_{\text{VBO}}$ and the known band gap of MgO, we conclude that the band offsets for the conduction bands are $>1$V for all interfaces considered.
This is different from the previous investigations of the interfaces of Co$_2$MnGe with GaAs or Ge, where DFT calculations found Fermi-level pinning  by gap states, or alignment of the valence bands, respectively\cite{Picozzi2003a}.

\begin{figure*}
\centering
\includegraphics[width=\textwidth]{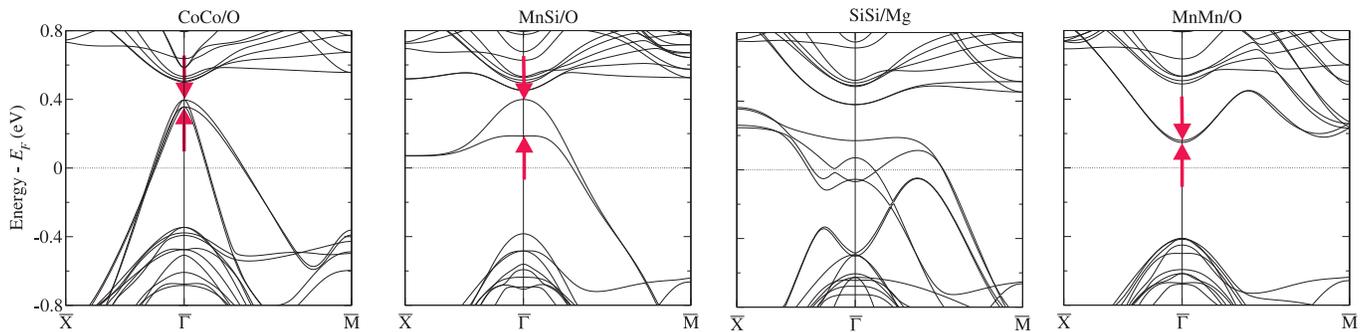}
\caption{Minority spin band structures of the TMR elements. Only for the MnMn/O interface half-metallicity is retained. In the other cases interface states appear at the Fermi energy. The splitting of interface states with even and odd symmetry due to tunneling is indicated by the arrows.}
\label{fig_mtj_band}
\end{figure*}

Finally, we discuss the role of the interface states for the electrical properties. 
In order to achieve a large TMR ratio, it is crucial that the current through the MgO barrier is minimal in the state of anti-parallel magnetization of the two electrodes. 
For perfect half-metallic electrodes, this current results from a sequence of spin-flipping and tunneling to or from an interface state. 
We make the (conservative) assumption that the latter process is rate-limiting (cf. the discussion in Ref.~\onlinecite{Mavropoulos2005}). To estimate the TMR ratio, one needs to take into account that the tunneling probability of carriers in the various bands may be vastly different depending on band symmetry. 
For MgO, it has been demonstrated that band states of $\Delta_1$ symmetry show a weak decay inside the MgO barrier, while transmission of states with different symmetry is exponentially suppressed.\cite{Mathon2001, Butler2001} 
For \cms, DFT calculations show that a conduction band in the majority spin channel with $\Delta_1$ symmetry exists.\cite{Miura2007} Hence, transmission of spin-majority  electrons will be easy for {\it parallel} magnetization of the TMR electrodes, which is one of the pre-requisites for achieving a high TMR ratio.
The crucial point, however, is 
the efficient suppression of currents for {\it anti-parallel} magnetization, which requires the tunneling rate into the minority interface states to be small.
From our calculations, we can estimate this rate from the energetic splitting between bands having even and odd symmetry with respect to the mirror plane inside the MgO barrier. 
In Fig. \ref{fig_mtj_band}, the minority spin bandstructures are shown. 
It is seen that the interface state at MnSi/O shows considerable splitting at $\overline \Gamma$, while the interface states at CoCo/O and MnMn/O show only a very small splitting of 41 meV and 11 meV, respectively (arrows in Fig.~\ref{fig_mtj_band}). As seen from Fig.~\ref{fig_mtj_dos_orbitale}, the interface band at MnSi/O consists of $3d$ orbitals {\em perpendicular} to the interface, 
which are invariant under symmetry operations in the interface plane, and thus match the $\Delta_1$ band of MgO; hence the large transmission. 
For the CoCo/O and MnMn/O interfaces, the interface bands formed by the in-plane $3d$ orbitals 
transform under various $\Sigma$-representations of the in-plane symmetry group; therefore  the coupling to the $\Delta_1$-type MgO band and the interface band splitting are small.
As expected for a splitting due to tunneling, it is  largest at the $\overline \Gamma$ point, since the tunneling probability is largest for vanishing parallel crystal momentum. 
Knowledge of the band structure puts us in position to speculate about the expected current-voltage characteristics of a TMR devices. 
For the MnMn/O interface, a bias voltage of $\sim 0.2$eV must be applied before electrons can be injected into the unoccupied interface band. 
For the CoCo/O interface, the splitting of the interface state, and hence the tunneling rate of spin-minority electrons, is tiny at $E_F$. Only after applying a bias voltage of $\sim0.35$~eV, 
raising $E_F$ in the spin-majority electrode, 
efficient tunneling into the spin-minority interface state near $\overline \Gamma$ will be energetically allowed. 
In TMR devices, the step-wise increase of the anti-parallel tunneling conductance as a function of bias voltage, as observed e.g. in Ref.~\onlinecite{Tsunegi2008} and \onlinecite{Chioncel2008},  
at low temperatures, 
may be an indication of the role of interface states in transport. Then, the achievable TMR ratio in present TMR devices would still be limited by the (improvable) interface quality, rather than by factors intrinsic to the Heusler alloy, as suggested by Ref.~\onlinecite{Chioncel2008}.

In summary, our DFT calculations indicate that TMR devices fabricated from \cms/MgO(100) heterostructures may form thermodynamically stable CoCo/O or MnSi/O interfaces, while the (most desirable) half-metallic MnMn/O interface may be prepared under non-equilibrium conditions. 
We conclude that interfaces made from a single atomic species (Co or Mn) enable a high TMR ratio, since the tunneling  of  the associated electronic interface states in the minority-spin channel through the MgO barrier is weak, which is a pre-requisite for a tolerably low  parasitic current in the anti-parallel magnetization of the TMR device.

\end{document}